# Race, Gender and Beauty: The Effect of Information Provision on Online Hiring Biases

**Weiwen Leung**
iamweiwenleung@gmail.com

**Zheng Zhang**
University of Rochester
Rochester, NY, United States
zzhang95@cs.rochester.edu

**Daviti Jibuti**
CERGE-EI
Prague, Czech Republic
djibuti@cerge-ei.cz

**Jinhao Zhao**
jinha14@tsinghua.org.cn

**Maximillian Klein**
max@notconfusing.com

**Casey Pierce**
University of Michigan
Ann Arbor, United States
cbspierc@umich.edu

**Lionel Robert**
University of Michigan
Ann Arbor, United States
lprobert@umich.edu

**Haiyi Zhu**
Carnegie Mellon University
Pittsburgh, United States
haiyiz@cs.cmu.edu

**ABSTRACT**
We conduct a study of hiring bias on a simulation platform where we ask Amazon MTurk participants to make hiring decisions for a mathematically intensive task. Our findings suggest hiring biases against Black workers and less attractive workers, and preferences towards Asian workers, female workers and more attractive workers. We also show that certain UI designs, including provision of candidates' information at the individual level and reducing the number of choices, can significantly reduce discrimination. However, provision of candidate's information at the subgroup level can increase discrimination. The results have practical implications for designing better online freelance marketplaces.

**Author Keywords**
discrimination; gig economy; hiring

**CCS Concepts**
•**Human-centered computing** → **Human computer interaction (HCI);** *Haptic devices;* User studies;

## INTRODUCTION
An increasing number of Americans are earning money through freelance jobs obtained through online platforms. Indeed, a report from Pew Research Center [29] indicates that 8% of Americans earn money from these "digital gigs". Websites that host these services facilitate supplemental income for some workers, and become a primary income source for others. The report also said that 14% of Black respondents and 11% of Latino respondents reported earning money on these platforms during the previous year, in contrast to 5% of White respondents. Among these non-White workers, 65% of them describe the income they earn from these platforms as "essential" or "important." Additionally, 55% of gig workers are female.

One important question is the extent to which different types of hiring biases exist on these platforms, both with respect to easily quantifiable characteristics such as race and gender, but also less easily quantifiable characteristics such as beauty.

Racial discrimination in offline hiring has been well documented, especially in the US. In particular, there is significant discrimination against African-Americans and Latinos in hiring, [4, 27]. Biases based on gender [5] and beauty [21] are also prevalent.

However, there are good reasons to suspect that online platforms may lessen or eliminate hiring biases. For example, Morton et al. [22] use observational data to show that while racial minorities pay 2% more for cars when purchasing them offline, this gap is much smaller for online purchases. They attribute this to the internet facilitating information search and removing cues present in offline negotiations. To the extent that such considerations are applicable, the internet may have a similar impact in the digital gig market.

Another underexplored question is whether user interface (UI) design factors can affect hiring biases. An answer would shed light on whether existing results on discrimination are largely a product of mutable factors such as UI design, or whether they are likely to generalize to other settings. More broadly, it would inform us if there are practical implications of knowing the factors that cause discrimination. Guryan et al. [12] note that this is an important unanswered question even in the economics literature on discrimination, which has existed for over 50 years.

To examine the prevalence of different types of biases in online hiring, as well as examine the effects of different design factors on biases, we conducted a study by setting up a task on Amazon MTurk in which we recruited 206 subjects to make hiring decisions in a platform simulating a website recruiting people for freelance jobs. We examine first and foremost how hiring rates are affected by gender, race, and beauty. We then examine whether the number of people displayed and/or performance information affect hiring decisions.

## LITERATURE REVIEW AND HYPOTHESES

**Biases and Discrimination in Hiring**

Since the seminal article by Bertrand and Mullainathan [4], the discrimination literature has seen explosive growth. Key areas of study have been racial and gender discrimination.

Edelman et al. [10] find that people with distinctively African-American names are 16 percent less likely to be accepted as guests on AirBnB compared to those with distinctively White names. Pope and Sydnor [25] study online loans on the peer-to-peer website Prosper.com, and find that loan listings with Blacks[1] in the attached picture are 25 to 35 percent less likely to receive funding than those of Whites with similar credit profiles. However, despite the higher average interest rates charged to Blacks, lenders making such loans earn a lower net return compared to loans made to whites with similar credit profiles because Blacks have higher relative default rates. There is also evidence consistent with such discrimination in online environments. For example, an observational study found that Black people tend to get more negative reviews than other races [13], which could harm their employment opportunities. To our knowledge, there has been less study of discrimination against other races such as Asians.

Gender discrimination has also been studied; whether or not females are discriminated against depends heavily on the task and context. For example, Bohnet et al. [5] find pro-female discrimination in hiring on language tasks and anti-female discrimination in mathematics tasks when candidates are evaluated one at a time. However, discrimination disappears when candidates are evaluated jointly. Coffman et al. [9] study gender discrimination when candidates are evaluated two at a time for male stereotyped tasks, and find discrimination when two candidates' prior performance are equal, but not when there is a candidate with a stronger prior performance. Finally, a field experiment on mathematics Stackexchange [6] found that low-reputation users with female usernames receive less upvotes for questions they post relative to those with male usernames. However, the direction of discrimination reverses at high reputation levels: those with female usernames receive *more* upvotes. The authors explain their findings could be due to people having incorrect belief about female math ability. Interestingly, there is no evidence for gender discrimination with regards to posted answers, and the authors attribute it to the decreased subjectivity over whether answers should be upvoted (as compared to questions). Gender discrimination can also vary over time; a 2017 study of LinkedIn data found that gender discrimination has decreased significantly over the past 10 years [30].

A small but growing literature examines how decision makers are affected by attractiveness. In a highly cited lab experiment, Mobius and Rosenblat [21] find a sizable beauty premium in hiring, as physically attractive workers are more confident and considered more able by employers, and are also thought to have better oral skills. Jenq et al. [17] study an online charitable microfinance website, and find that borrowers who are more attractive receive funding more quickly.

In this paper, we explore the extent to which different forms of hiring biases based on gender, race, and attractiveness can manifest themselves in a online freelancer marketplace, and then examine the effect of UI design on hiring biases.

We focus on math as our task domain because race-based and gender-based stereotypes on math are well-documented in the literature [11]. Furthermore, multiple sources indicate gender and racial gaps in SAT math scores that have persisted over time. In particular, males outperform females[2], Asians outperform Whites, and Whites outperform both Blacks and Latinos[3].

Based on these, we formulate H1 to H5. Note that the hypotheses are formulated under the assumption that in each hypothesis, workers from each subgroup are on average equally-qualified[4] from the employer's perspective (i.e. what the employer can observe). We design our experimental trials such that this assumption holds.

**H1.** *Females will be hired less frequently than males.*

**H2.** *Asians will be hired more frequently than Whites.*

**H3.** *Whites will be hired more frequently than Blacks.*

**H4.** *Whites will be hired more frequently than Latinos.*

**H5.** *A more beautiful person will be hired more often than a less beautiful person.*[5]

---

[1] We use "Black" instead of "African-American" to be consistent with the original study. In the rest of the paper, we use terminology that is consistent with the underlying sources as far as possible.

[2] http://www.aei.org/publication/2016-sat-test-results-confirm-pattern-thats-persisted-for-45-years-high-school-boys-are-better-at-math-than-girls/, https://www.fairtest.org/sat-act-gender-gaps

[3] https://www.brookings.edu/research/race-gaps-in-sat-scores-highlight-inequality-and-hinder-upward-mobility/, https://www.insidehighered.com/news/2017/09/27/scores-new-sat-show-large-gaps-race-and-ethnicity

[4] For example, candidates without any prior observable performance are equally-qualified. Candidates with the same observable performance are also equally-qualified.

[5] We are not aware of any data that examines the correlation between beauty and math test scores. However, H5 is based on the studies we cited [17, 21] which found more favorable outcomes for beautiful people.

**User-Interface Design**

There are many ways in which user interface design can affect behavior. One important way is through the provision of information. For example, certain subgroups may be less likely to be hired as they are perceived to be less productive than other subgroups, a phenomenon known as statistical discrimination. However, the provision of information on individuals' performance on previous tasks can reduce statistical discrimination [3].

In contrast, the effect of information on how certain subgroups performed can either reduce or increase hiring disparities across different subgroups. For example, if subgroup A is hired more often than subgroup B, but information reveals that both subgroups are equally productive, then information about subgroup performance should reduce the gap. In contrast, if both subgroups are hired equally often, but information reveals that subgroup A is more productive, then subgroup performance information should result in workers from subgroup A being hired more often.

Based on these, we propose H6 and H7.

**H6.** *Provision of information at the individual level (how the candidate did in previous tasks) can reduce hiring bias.*

**H7.** *Provision of information at the subgroup level (how the candidate's subgroup did in previous tasks) moves hiring biases in the direction of the productivity difference across different subgroups.*

In our experiment, we examine how the provision of information at the individual level (how the candidate did in previous tasks) and subgroup level (how one's subgroup did in previous tasks) affect hiring decisions.

Another way that UI changes can affect decision making is by altering the choice environment by using behavioral "nudges" [31]. For example, Lee et al. [20] find that behavioral economics persuasion techniques such as having default options can lead to people making healthier food choices.

One well-known nudge is to vary the number of options to choose from (i.e. the size of the choice set). A famous study showed that people are much more likely to buy jam when faced with 6 varieties than when faced with 24 varieties [16], a phenomenon known as "choice overload". While we know that increasing the number of options makes one less likely to make a choice [8], what is less well known is the effect on *which* choice is made. Our study contributes to this literature by providing more insight into how choice overload affects which choice is made, with a focus on equity concerns.

In the context of online hiring, we propose that the size of the choice set can influence hiring biases. One natural hypothesis may be that increasing the number of candidates for hire may lead to people use heuristics - gender-based or race-based stereotypes. Indeed, under Kahneman's dual system framework, people are more likely to use heuristics when overloaded with information [18], and one of the few studies examining the impact of choice set size on which choice is made found that people tended to go with easy-to-understand (e.g. less risky) options when the choice set expanded [15]. We hypothesize that people are more likely to use heuristics that will accentuate existing biases (e.g. those based on stereotypes) when faced with a larger choice set, and hence formulate our eighth hypothesis as below.

**H8.** *Increasing the number of candidates to choose from can increase hiring bias.*

**EXPERIMENT DESIGN**

We designed an experiment where participants were told they would be making hiring decisions for a mathematically intensive task. We told participants[6] that potential employees had completed two sets of mathematical questions, one easy set and one difficult set (Round 1 and Round 2 respectively). Both sets had five questions each. We showed participants example questions from both sets. The easy questions were similar in difficulty to easy SAT questions, and the difficult questions were similar in difficulty to difficult SAT questions.

We designed our experiment around mathematically intensive tasks for several reasons. First, clear stereotypes exist, at least with regards to gender [11]. Second, the discrimination literature often uses mathematically intensive tasks as a subject of study [5, 6, 9]. Finally, many gig work tasks involve the use of mathematics: a search of sites such as Fiverr and Upwork reveal thousands of math-related tasks.

All participants were told there would be twelve hiring rounds, and they would make one hiring decision in each round. Participants were told they would see the photos of potential employees. A third of participants were told they would see two potential employees in each round, while another third were told they would see four, and the remaining third were told they would see eight. Figure 1 shows a screenshot of the two potential employees condition.

(In reality, the photos of potential employees were drawn from the Chicago Face Database, so that we could obtain measures of perceptions of race, gender, and attractiveness. However, we did ensure that the gender of the photo corresponded to the gender of the potential employee we in fact hired from MTurk to answer SAT-level math questions. We did not take photos of the people we had actually hired to solve mathematical questions because MTurk does not allow us to take or request photos from MTurkers.)

To encourage participants to take hiring decisions seriously, participants were told that after they made all hiring decisions, one of their hires would be randomly selected, and they would be given a bonus of $1 for every question their person they hired on a randomly selected round had correctly solved on the difficult set[7].

A third of all participants saw the number of questions that potential employees correctly solved on the easy set (Figure 1 shows a screenshot of a trial with such information), while a third of all participants saw the performance distribution by gender of questions that potential employees correctly solved on the easy set before they made any hiring decisions (screenshot available in Figure 2). The remaining third of participants did not see either.

---

[6]We recruited U.S. MTurkers who had completed at least 500 tasks and an acceptance rate of at least 97%.
[7]There were a total of five questions, so the maximum bonus payout was $5

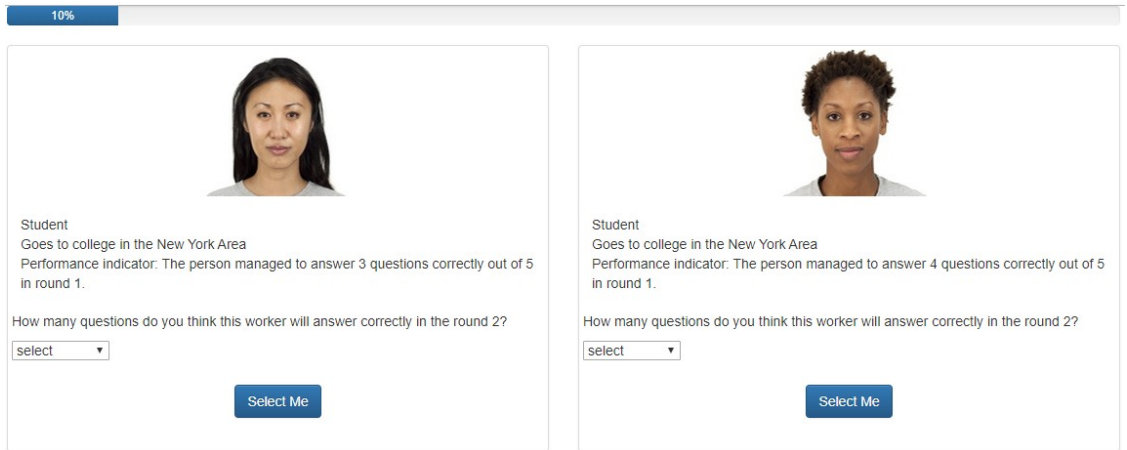

Figure 1. Screenshot of two worker condition with performance on easy questions displayed.

Because the photos shown to participants were chosen randomly from the Chicago Face Database (except with respect to gender), the expected past performance of potential employees of each race was equal. Likewise, the expected past performance of potential employees was unaffected by their beauty. In addition, experimental trials were designed so that the average past performance of workers (that were displayed to participants) across genders would be exactly equal[8]. Therefore, if the race, gender, and beauty of a potential employee were immaterial to our participants, we should discover that these factors had no effect on hiring.

All participants were given comprehension questions to make sure they understood the nature of the experiment (including that their payout would depend on the hiring decisions they made), and had to answer the comprehension questions correctly before they could proceed with hiring decisions.

Observe from Figure 1 that before each hiring decision, we asked participants to predict the number of difficult questions each worker would answer correctly. This technique is known in the discrimination literature as "belief elicitation" (see e.g. [9]) and is used to examine whether discrimination (if present) is due to people's beliefs about the productivity of different subgroups.

**DATA ANALYSIS METHODOLOGY**
We use discrete choice modelling to analyze our data. A discrete choice model is a form of agent-based model that is often used in economics [23], marketing [2], transportation [1], and public health [19], among other fields.

In a discrete choice model, a decision maker chooses between different alternatives (e.g. products, healthcare options, transportation options, or in our case, potential employees). The decision maker computes the value of each alternative as the function of that alternative's characteristics.

Suppose that a decision maker is considering $N$ alternatives (in our case, potential employees). In our case, a decision maker might compute the value of potential employee $i$ ($i \in \{1, 2, ..., N\}$) follows:

$Value_i = \beta_0 + \beta_1 Female_i + \beta_2 Asian_i + \beta_3 Black_i + \beta_4 Latino_i + \beta_5 Attractiveness_i$

where $Female_i$, $Asian_i$, $Black_i$, $Latino_i$ are variables indicating the gender and race of the worker. $Attractiveness_i$ is a continuous variable measuring the attractiveness of the worker. We did not ask our participants to evaluate the attractiveness, race, or gender of each potential employee (and doing so would be time consuming and interfere with participants' decisions[9]). Instead, we proxied these variables by using their values from the corresponding photo in the Chicago Face Database, which was based on the results of a survey on the proportion of people who thought the person in the photo was female, Asian, Black[10], or Latino, as well as the average attractiveness rating of the photo. This introduces measurement error, but classical measurement error biases our coefficient estimates towards zero, making it harder for us to find effects that in fact exist [32].

Decision makers want to choose the option with the highest value. However, decision makers measure value with error e.g. because of errors in perception, errors in computation, or due to randomness in taste. The chance that they will choose a particular option is therefore a probabilistic function that increases as the value of that particular option increases, and decreases as the value of alternative options increase. The exact mathematical equations governing our model can be found in the Appendix.

---

[8] Details of our experimental trials can be found in the Appendix.

[9] However, we did ask participants to predict the number of difficult questions each participant would answer correctly, because such information was valuable and could not be proxied by data from other sources.

[10] To be consistent with Chicago Face Database terminology, we use "Black" and not "African-American"

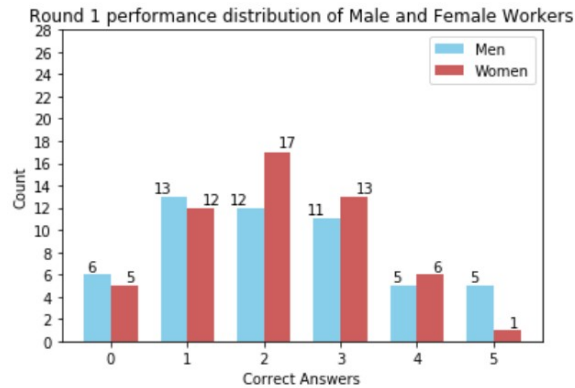

**Figure 2.** Distribution of performance by gender, which was shown to randomly selected participants before they made their hiring decisions, along with a short explanation of how to interpret the distributions presented.

The discrete choice model has several desirable properties. Perhaps most importantly, it takes into account that decision makers take into account the *relative* value of each alternative when making a decision. For example, an option with a value of 10 would likely be chosen if there was only one alternative option with a value of 1, but not if the alternative option had a value of 100[11]. The model also flexibly adjusts to the fact that the probability of choosing any alternative decreases when more choices are available, which is important for our case since we vary the number of potential employees.

We use maximum likelihood estimation (MLE) to estimate the parameters ($\beta$'s) of our model. We cluster standard errors by participant because each participant makes multiple decisions[12].

Estimating the model using our entire sample allows us to estimate the overall effects of race, gender, and attractiveness. To evaluate the effects of these variables under each experimental manipulation, we estimate the model using only data from participants who were exposed to that experimental manipulation.

Recall that in the discrete choice model, our decision makers estimate value. Thus our coefficient estimates should be interpreted as the marginal (additional) value of a given attribute. The effect on the probability of hiring can be computed through the use of odds ratios[13]. Note that since the probability of hiring is monotonically increasing in value, positive coefficient estimates always indicate a positive effect of a given attribute on the probability of hire.

Especially because our experiment involves manipulating race, gender, and beauty through the use of photos, our coefficient estimates should largely be interpreted as descriptive (correlational) rather than causal. Indeed, when our participants make hiring decisions, they may consider other factors besides race, gender and attractiveness. For example, suppose that people of a certain race are perceived as less trustworthy. Then what our model attributes to race may actually be *caused by* perceived trustworthiness. However, because our main interest lies in understanding the extent to which certain subgroups of workers face discrimination (e.g. African-Americans), regardless of the underlying cause, our coefficient estimates will actually capture the desired effect. Additionally, in our additional analyses/robustness check section, we examine whether our results change when we add in other variables regarding the appearance of the person in the photo (e.g. trustworthiness).

Recall also that our experiment was designed such that the workers from the different subgroups of interest were in fact on average observably equally-qualified. Hence, we can make claims regarding observably equally-qualified workers without controlling for previous performance. We do not control for previous performance because we do not display previous performance of workers to half of our participants.

Before we discuss the results, we briefly note two limitations of our methodology. First, even though attractiveness was measured by independent coders engaged by the Chicago Face Database, notions of attractiveness may reflect Western concepts. Second, while our data covers male and female genders well, we may not be able to generalize to other genders.

**RESULTS**

**Overview of the Findings**
Our main findings include:

- Results suggested hiring biases against Black candidates, and towards Asian candidates and female candidates.

---

[11] A standard linear/logistic regression only makes use of a given alternative's characteristics, and it would be impossible or extremely difficult to replicate the flexibility of the discrete choice model by adding control variables especially since the size of the choice set varies across participants.

[12] The usual formula for calculating standard errors is only valid when one participant makes one hiring decision. Making multiple decisions could introduce serial correlation. Using clustered standard errors allows us to adjust for such serial correlation by making it harder for us reject the null hypothesis relative to using the usual standard errors, as clustered standard errors are usually bigger [7].

[13] The odds ratio of a coefficient estimate of $X$ is $e^X$. Specifically, if a coefficient estimate of $X$ indicates that a one unit increase in the explanatory variable is associated with a $e^X - 1$ increase in the probability of hire, holding other explanatory variables constant. For example, $X = 0.1$ would correspond to roughly 10.5% increase.

Specifically, Black candidates are hired 16% less compared to White candidates; Asian candidates are hired 23% more compared to White candidates; female candidates are hired 61% more than male candidates; there are no significant difference between White candidates and Latino candidates. H2 and H3 are supported, while H1 and H4 are not supported.

- Results suggested hiring biases toward more attractive candidates. One standard deviation increase in the attractiveness increased hiring chances by around 10%. H5 is supported.

- Provision of information at the individual level (how the candidate did in previous task) erased the differences between White candidates, Asian candidates and Black candidates, and the difference between attractive candidates versus non-attractive candidates, though it further increased the hiring chances of female candidates. H6 is largely or completely supported (depending on how one interprets the further increase in hiring chance for female candidates).

- Provision of information at the subgroup level (how the candidate's subgroup did in previous task) by gender did not decrease the hiring chances of female candidates. However, it reduced the hiring chance for Black candidates, and increased the hiring chance for Asian candidates. We subsequently discuss our interpretation of these results.

- Increasing the number of candidates to choose from reduced the chance that Black candidates would be hired. H8 is supported.

**Descriptive Statistics**

Table 1 gives descriptive statistics associated with potential employees by gender. The first row indicates performance in Round 1 (easy SAT math questions) by gender of the candidates that we hired to answer math questions. We see that males perform better in the Round 1 compared to females, however, the difference is insignificant. As the table shows, females answered, on average, 2.04 questions correctly, while males managed to answer 2.21. Overall in the Round 1 workers had 5 questions to answer.

The second row indicates the mean predicted score of candidates[14] that were given by participants in our experiment. Participants expected females to perform significantly better in Round 2 (harder SAT math questions). Females were expected to outperform males. Many explanations for this difference are possible, but one possible explanation behind the higher prediction for female workers is that participants may have thought that gender discrimination was the topic of study and tried to counteract any implicit biases they held. Note also that predicted scores could have been higher for harder SAT math questions than actual scores for easy SAT questions because most participants did not see the distribution of scores for easy SAT questions.

The third row indicates the mean attractiveness score of photos that appeared in our experiment (by gender) as given by coders in the Chicago Face Database. Females were perceived to be more attractive than males.

We now focus on participants who made hiring decisions (i.e. employers). They came from a wide variety of backgrounds. For example, they came from 34 U.S. states; the three states which contributed the most number of participants had 18%, 12%, and 8% of the subject pool, and all other states each contributed 4% or less. 42% of the participants are females and are rest males. In terms of their highest educational level, 3% of our sample have a high school diploma or lower and 17% have either some college or a 2 year college degree. 45% of our subjects have 4 years of college degree, 29% have a masters degree and 6% have a professional degree. Although the sample skews towards the more educated, one might expect that the more educated are more likely to hire people in the gig economy due to higher income. Unfortunately, we did not collect data on race or mathematical ability[15].

When asked explicitly whether math was associated with a particular gender, subjects tended to associate math with males (rather than females). Only 17% of our sample somewhat associate math with females, with 26% do not associate math with any gender and remaining 43% associate it with males. On the other hand, approximately equal numbers of subjects associate liberal arts with females (35%) and males (32%), while a third of participants (33%) think itâĂŹs not related to any specific gender. Note that demographics and opinions were only collected at the end of the study.

| | Female | Male | $p-value$ |
|---|---|---|---|
| Round 1 performance | 2.04 | 2.21 | 0.49 |
| | (0.16) | (0.20) | |
| Prediction | 2.73 | 2.66 | 0.006 |
| | (0.02) | (0.02) | |
| Attractiveness score | 3.44 | 3.08 | 0.000 |
| | (0.01) | (0.01) | |

**Table 1.** *Notes:* Summary statistics by *Gender*. Standard errors in parenthesis beneath mean estimates. The last columns shows p-values of the hypothesis of equal means across groups.

In terms of race of the photos we took from the Chicago Face Database, almost a third of workers were White (31%) followed by African American (28%), Asian (22%) and Hispanics (19%).[16]

**Discrete Choice Model: Results with full sample**

Estimating our table on the full sample indicates that attractive candidates are valued more and hence hired more often than observably equally-qualified unattractive candidates, as evidenced by the positive and statistically significant coefficient of $Attractive_i$. Analogously, female candidates are valued more and hence hired more often compared to observably equally-qualified male candidates, and Asian candidates are hired more often compared to observably equally-qualified White candidates (male and White are omitted categories; we do not mention the omitted categories from now on for brevity). However, Black candidates are valued less and hence hired less often relative to observably equally-qualified White candidates (see leftmost column of Table 2). (**Note:** We omit "observably equally-qualified" for brevity from now on. We'll also use "hired more often" instead of "valued more and hence hired more often" from now on, and likewise "hired less often" means "valued less and hence hired less often".)

---

[14] Candidates that appeared to participants in our experiment. Recall that while we did hire MTurkers to solve easy and difficult math questions, we were not allowed to take photos of them, or ask them to supply photos. So we took photos of people from the Chicago Face Database, and matched it with the the workers we hired based on gender information.

[15] It might be useful for future work to examine the influence of mathematical ability on discrimination.

[16] We define race in line with the Chicago Face Database definition.

In sum, H2, H3 and H5 are supported, but H1 and H3 are not supported (we in fact observe the reverse of H1). As we discussed earlier, one possible explanation behind the higher hiring rates for female candidates compared to male candidates is that participants tried to counteract any implicit biases they held against female candidates.

**Showing prior performance**
When we do not show any performance information on easy mathematics questions, all explanatory variables that were significant in the full sample (i.e. in the analysis immediately preceding this) remain significant and have the same sign, with the exception of $Black_i$. The results indicate that Black candidates are not chosen at a different rate compared to White candidates (see second column of Table 2, which has heading "*None*").

When we show candidates' individual performance on easy mathematics questions, attractiveness and race have no statistically significant effect on hiring (see "individual" column of Table 2). Female candidates are still hired more often, and in fact the coefficient estimate of *Female* increases compared to when no information about prior performance is displayed. We conclude that H6 is largely or completely supported, depending on how one interprets the further increase in likelihood of female candidates being hired.

When we show the distribution of candidates' performance on easy questions by gender, the chances of female candidates being hired did not decrease[17], even though women actually performed slightly worse than men on the easy SAT-level math questions (relative to no information on prior performance). Although a more comprehensive test would have examined the impact of information on performance by other factors such as race (we only displayed subgroup information by gender to maximize statistical power), the available evidence does not support H7. However, we note that information on subgroup performance by gender reduced the hiring chances of Black candidates, while increasing the hiring chances of Asian candidates. Therefore, displaying performance by gender could have had behavioral effects e.g. trigger subconscious stereotypes about race.

**Number of candidates**
Among participants who were asked to choose between two candidates at a time, we find that attractive candidates and fe-

[17]and actually increases slightly, though the difference is not statistically significant at conventional levels

|  | Full sample | Prior performance shown | | |
|---|---|---|---|---|
|  |  | *None* | *Individual* | *Subgroup* |
| Attractiveness | 0.14*** | 0.24*** | 0.03 | 0.16*** |
|  | (0.03) | (0.06) | (0.06) | (0.06) |
| Female prop | 0.48*** | 0.36*** | 0.67*** | 0.42*** |
|  | (0.06) | (0.09) | (0.10) | (0.10) |
| Asian prop | 0.21** | 0.30** | -0.07 | 0.36** |
|  | (0.08) | (0.15) | (0.15) | (0.14) |
| Black prop | -0.17** | 0.02 | -0.18 | -0.38*** |
|  | (0.07) | (0.12) | (0.12) | (0.13) |
| Latino prop | -0.12 | 0.20 | -0.21 | -0.37 |
|  | (0.12) | (0.20) | (0.22) | (0.23) |
| Pseudo $R^2$ | 0.025 | 0.026 | 0.032 | 0.031 |

**Table 2.** *Notes:* **The table shows estimation results of the Discrete Choice Model. Standard errors are clustered at the subject level.** $*p < 0.1; **p < 0.05; ***p < 0.01$.

male candidates are hired more often. The variables indicating race are not statistically significant (see Table 3, column "2").

Participants who were asked to choose between four candidates at a time chose attractive candidates, female candidates and Asian candidates more often (and the effect is statistically significant at conventional levels). Black candidates are chosen less often, but the difference is not statistically significant at conventional levels (see Table 3, column "4").

Finally, participants who were asked to choose between eight candidates at a time chose attractive candidates and female candidates more often, and this difference is statistically significant. Black candidates were chosen less often. Asian candidates were chosen more often, though the difference is not significant at conventional levels (see Table 3, column "8").

|  | Worker condition | | |
|---|---|---|---|
|  | *2* | *4* | *8* |
| Attractiveness | 0.14** | 0.13** | 0.15*** |
|  | (0.06) | (0.06) | (0.06) |
| Female prop | 0.55*** | 0.30*** | 0.62*** |
|  | (0.10) | (0.09) | (0.10) |
| Asian prop | 0.10 | 0.35** | 0.15 |
|  | (0.15) | (0.14) | (0.14) |
| Black prop | -0.08 | -0.13 | -0.29** |
|  | (0.13) | (0.12) | (0.13) |
| Latino prop | -0.07 | 0.02 | -0.23 |
|  | (0.23) | (0.20) | (0.22) |
| Pseudo $R^2$ | 0.035 | 0.017 | 0.031 |

**Table 3.** *Notes:* **The table shows estimation results of the Discrete Choice Model. Standard errors are clustered at the subject level.** $*p < 0.1; **p < 0.05; ***p < 0.01$.

**Additional analyses**
To examine whether the effects of attractiveness differ by the gender of the candidate, we add the interaction of *Female* and *Attractiveness* to our main specification. As the estimation result shows, the coefficient of the interaction term is not statistically significant, suggesting that increasing attractiveness has the same effect for male and female candidates (see Table 4, column 1).

When we add the predicted number of difficult math questions that candidates got right to our main specification[18], we find that predicted score is positively and strongly correlated with the hiring decision. However, even after controlling for predicted score, female and attractive candidates are still hired more often. Although the coefficients of race still have their expected signs, they become statistically insignificant at conventional levels (see Table 4, column 2). It appears that, at least for race, differences in hiring races can be explained by differences in predicted performance.

|  | (1) | (2) |
|---|---|---|
| Attractiveness | 0.08* | 0.08** |
|  | (0.05) | (0.04) |
| Female prop | 0.48*** | 0.54*** |
|  | (0.06) | (0.06) |
| Female*Attractiveness | 0.05 |  |
|  | (0.06) |  |
| Asian prop | 0.20** | 0.09 |
|  | (0.09) | (0.09) |
| Black prop | -0.17** | -0.08 |
|  | (0.07) | (0.08) |
| Latino prop | -0.12 | -0.09 |
|  | (0.12) | (0.14) |
| Prediction |  | 1.07*** |
|  |  | (0.04) |
| $PseudoR^2$ | 0.025 | 0.24 |

**Table 4.** *Notes:* The table shows estimation results of the Discrete Choice Model. Standard errors are clustered at the subject level. $*p < 0.1; **p < 0.05; ***p < 0.01$. Column 1 adds "female" and "Attractiveness" interaction to the main specification, while column 2 adds "Prediction" as an additional independent variable.

**Are effects driven by other characteristics?**
To find possible underlying mechanisms, as well as to check whether other facial expression characteristics affect hiring decisions, we included those characteristics (as rated by independent coders hired by the Chicago Face Database) as explanatory variables. Being perceived as angry significantly reduces workers' chance of being employed (see Table 5, column 1), while appearing happy increases that probability, though the effect is statistically insignificant (see Table 5, column 2). When *Angry* or *Happy* are added as control variables, all explanatory variables that were statistically significant in the original specification remain significant and have the same sign. In column 3, we examine what happens when perceived masculinity is added to the model. Workers that are perceived to be masculine are less likely to be hired, and in this model, the coefficient estimate of gender is no longer statistically significant at conventional levels. Column 4 examines what

[18]Recall that we got participants to predict the number of difficult questions candidates answered correctly before making their hiring decision

happens when *Feminine* is added to the model. People that appear feminine are more likely to be hired, and once *Feminine* is added, the effects of attractiveness and gender become much smaller in magnitude and statistically insignificant (see Table 5, column 4). This suggests that appearing feminine (or masculine) may be an underlying mechanism for some of our results.

|  | (1) | (2) | (3) | (4) |
|---|---|---|---|---|
| Attractiveness | 0.11*** | 0.13*** | 0.12*** | 0.05 |
|  | (0.04) | (0.04) | (0.04) | (0.05) |
| Female prop | 0.50*** | 0.49*** | 0.16 | -0.00 |
|  | (0.06) | (0.06) | (0.16) | (0.20) |
| Asian prop | 0.18** | 0.20** | 0.17* | 0.19** |
|  | (0.09) | (0.09) | (0.09) | (0.09) |
| Black prop | -0.18** | -0.18** | -0.13* | -0.14* |
|  | (0.07) | (0.07) | (0.07) | (0.07) |
| Latino prop | -0.11 | -0.11 | -0.11 | -0.09 |
|  | (0.12) | (0.12) | (0.12) | (0.12) |
| Angry | -0.11** |  |  |  |
|  | (0.04) |  |  |  |
| Happy |  | 0.04 |  |  |
|  |  | (0.04) |  |  |
| Masculine |  |  | -0.13** |  |
|  |  |  | (0.06) |  |
| Feminine |  |  |  | 0.18** |
|  |  |  |  | (0.07) |
| $PseudoR^2$ | 0.026 | 0.025 | 0.026 | 0.026 |

**Table 5.** *Notes:* The table shows estimation results of the Discrete Choice Model. Standard errors are clustered at the subject level. $*p < 0.1; **p < 0.05; ***p < 0.01$. Various characteristics from the Chicago Face Database are added to the main specification in this table.

Further examining other facial expressions shows that "Dominant" and Threatening" face significantly reduces hiring probability and "Trustworthy" face does not have significant impact on the result (see Table 6, columns 1-3). None of these variables affect the significance of the other variables.

We also conduct additional robustness checks by removing outliers (e.g. people that responded too quickly or slowly). Results are in the Appendix.

**EFFECT SIZES**
Recall that we can use odds ratios to compute effect sizes. For example, in the main sample, we find that Blacks are 16 percent less likely to be chosen than Whites ($e^{-0.17} - 1 = -0.16$). The effect size is smaller than in Bertrand and Mullainathan [4], who find that African-Americans are 50 percent less likely to receive interview callbacks than Whites. However, the magnitude of discrimination still appears to be sizeable.

The effects of our experimental manipulations are economically meaningful as well. For example, in the two worker condition, Blacks are chosen around 7 percent less than Whites, but in the eight worker condition they are chosen around 25 percent less than Whites.

We give tables with odds ratios in the Appendix.

|                | (1)      | (2)      | (3)      |
|----------------|----------|----------|----------|
| Attractiveness | 0.14***  | 0.11**   | 0.11***  |
|                | (0.03)   | (0.04)   | (0.04)   |
| Female prop    | 0.42***  | 0.48***  | 0.46***  |
|                | (0.06)   | (0.06)   | (0.06)   |
| Asian prop     | 0.16*    | 0.17**   | 0.15*    |
|                | (0.09)   | (0.09)   | (0.09)   |
| Black prop     | -0.14*   | -0.19*** | -0.17**  |
|                | (0.07)   | (0.07)   | (0.07)   |
| Latino prop    | -0.06    | -0.10    | -0.11    |
|                | (0.13)   | (0.13)   | (0.12)   |
| Dominant       | -0.13*** |          |          |
|                | (0.05)   |          |          |
| Trustworthy    |          | 0.13     |          |
|                |          | (0.09)   |          |
| Threatening    |          |          | -0.14**  |
|                |          |          | (0.05)   |
| $PseudoR^2$    | 0.026    | 0.026    | 0.026    |

Table 6. *Notes:* The table shows estimation results of the Discrete Choice Model. Standard errors are clustered at the subject level. $*p < 0.1; **p < 0.05; ***p < 0.01$. **Various characteristics from the Chicago Face Database are added to the main specification in this table.**

## POTENTIAL INTERPRETATIONS

Our data do not allow us to pinpoint the underlying mechanism. However, we give an explanation that is consistent with the unexpected bias in favor of females, and the expected bias we found with regards to attractiveness and race.

This explanation is based on the notion that awareness can reduce one's subconscious biases. Racial bias in professional basketball referees persisted even after a study showed such bias [26], but disappeared after extensive media coverage of that study, suggesting that awareness reduced such bias [24]. Making crowdworkers aware of their own biases reduced their own biases [14], and academic promotion committees in scientific fields do not promote more men over women when they believe that gender bias exists [28].

It could be that participants thought that gender bias was the purpose of this study (being an often mentioned topic with regards to mathematical performance) and tried to correct for this bias, but were overzealous in correcting for it. However, were not aware of their subconscious racial and attractiveness biases in mathematics (perhaps because disparities by race and attractiveness in mathematics are less often mentioned) and did not correct for it.

Regarding the effects of our manipulations on user interfaces, we speculate that showing prior performance at individual level may have resulted in participants' attention diverted towards participants' past performance, hence the effects of all other characteristics disappeared, except for the most salient characteristic (gender). Analogous explanations for our other results may also be possible. For example, displaying information on performance across genders could have made participants more subconscious about other possible groupings of potential employees (e.g. race) and associated stereotypes.

We emphasize that future research should examine the validity of this explanation.

## IMPLICATIONS

The gig economy has many stakeholders, and each stakeholder can have multiple objectives (e.g. efficiency, equity). Here, we take the viewpoint of an online administrator concerned with equity.

One implication that stands out is that choice overload can negatively affect certain subgroups. Indeed, there was much less evidence of racial discrimination in the two worker condition than in the four or eight worker condition. If our results generalize, designers of the online freelance platforms should consider displaying candidates in a way that is less likely to trigger such choice overload. One possible technique that deserves further study is to limit the number of candidates displayed on each page.

A second implication is that designers should be careful in providing information designed to assist hiring decisions. Recall that the relationship between the amount of information and discrimination was not monotonic; there was some discrimination when no information on past performance was provided, the most discrimination when information on performance by subgroup (gender) was provided, and the least discrimination when individual level performance was provided. Since an intuitive explanation is that subgroup information could have increased discrimination by reminding people to consider a person's subgroup, designers of online freelance platforms should gather feedback before implementing significant UI changes, think carefully about equity concerns, and continually monitor key metrics even after changes are implemented to make sure that subgroups are not unnecessarily adversely affected.

The third, and potentially the most important, implication is that designers can consider making people aware of their subconscious biases. If our explanation that the lack of hiring bias against females was due to people being aware of this particular implicit bias is verified by future research, then online administrators can explore methods of making people aware of their biases.

## CONCLUSION

We ran an MTurk experiment where we asked participants to make hiring decisions for a mathematically intensive task. We unexpectedly find that our participants hire females more often than males. However, racial discrimination occurs largely as expected: Blacks are hired less often than Whites and Asians are hired more often than Whites. Also, attractive candidates are hired more often than less attractive candidates. Moreover, racial discrimination increases as the number of workers a participant can choose from increases. Finally, the relationship between discrimination and information provided to assist hiring decisions is non-monotonic in the amount of information provided.

The immediate takeaway is that since UI designs (reducing the number of choices and showing information of candidates at individual level) can reduce hiring biases, designers of online freelance platforms can do much to reduce hiring biases. Despite the limitations of our study in pinpointing the exact underlying mechanisms, our findings also serve as a call for further research in this area to determine under what contexts biases in hiring manifest themselves.

Our paper contributes to several literatures. Our finding that UI factors can affect discrimination is relevant to the human-computer interaction literature as well as the discrimination literature. By illustrating the use of a discrete choice model to measure discrimination, we also highlight to the HCI community how agent-based modelling can be used to estimate the value of different characteristics in situations where decision makers have to choose between varying numbers of alternatives, as well as alternatives that vary across decisions. We also contribute to the choice overload literature by verifying that choice overload can affect employment decisions, as well as by illustrating how choice overload can affect equity concerns.

We mentioned several limitations of our study previously at different points in the paper, but would like to mention a few more. One key limitation is generalizability: our study involved hiring people for mathematically intensive tasks. The kinds of discrimination that appear, as well as the methods of reducing such discrimination, may be different if the nature of the task were changed, particularly if the study were conducted in a field setting. Nonetheless, it is our belief that with persistent study and effort, it is possible to reduce discrimination in many areas, and our paper shows the potential of UI design to decrease discrimination.

Also, our findings may not generalize to settings without photos, such as Amazon MTurk[19]. That said, many online platforms use photos in their worker profiles, such as TaskRabbit, Upwork and Fiverr (to name a few). Even non-gig work marketplaces such as AirBnB, Uber, and Lyft use photos in their worker profiles (and racial discrimination based on photos has been documented in all three of them). Finally, the use of photos in offline resumes is common in European countries such as Germany, as well as China and Japan. Therefore, while it would be useful for future work to examine a setting without photos, we would argue that at the time of writing, an experiment that uses photos is at least as important (if not more important) than an experiment that does not.

### ACKNOWLEDGMENTS
We are grateful to Christine Exley, Max Harper, Loren Terveen, Brent Hecht, Teng Ye, and Jacob Thebault-Spieker for their help at different stages of the project. This work was supported by the National Science Foundation, under grant IIS-2001851, grant IIS-2000782, and grant IIS-1939606.

### NOTE
Weiwen Leung and Zheng Zhang are co-first-authors.

---

[19]It also will not generalize to MTurk because on MTurk, employers (or more precisely, requesters) do not choose workers. Rather, employers set criteria, and anyone who meets them can start the task.